\documentclass[runningheads]{llncs}
\usepackage{graphicx}
\usepackage{caption}

\usepackage{amsmath}
\usepackage{amsthm}
\usepackage{graphicx,psfrag,epsf}
\usepackage{enumerate}
\usepackage[square,sort,comma,numbers]{natbib}
\setcitestyle{square}
\usepackage{url}
\usepackage{amssymb}
\usepackage{color,soul}
\DeclareMathOperator*{\argmax}{arg\,max}
\DeclareMathOperator*{\argmin}{arg\,min}

\usepackage{algorithm}
\usepackage{algpseudocode}

\newcommand{\opt}{*}
\newcommand{\dd}{\mathop{}\! \mathrm{d}}
\newcommand{\appropto}{\mathrel{\vcenter{
  \offinterlineskip\halign{\hfil$##$\cr
    \propto\cr\noalign{\kern2pt}\sim\cr\noalign{\kern-2pt}}}}}
\usepackage{tikz}
\usetikzlibrary{arrows,automata,backgrounds,fit,patterns,petri,shadows,shapes}

\pgfdeclarepatternformonly{stripes}{\pgfpoint{0cm}{0cm}}{\pgfpoint{1cm}{1cm}}{\pgfpoint{1cm}{1cm}}{
  \foreach \i in {0.125, 0.375,...,0.875}{
    \pgfpathmoveto{\pgfpoint{\i cm}{0cm}}
    \pgfpathlineto{\pgfpoint{1cm}{1cm - \i cm}}
    \pgfpathlineto{\pgfpoint{1cm}{1cm - \i cm + 0.125cm}}
    \pgfpathlineto{\pgfpoint{\i cm - 0.125cm}{0cm}}
    \pgfpathclose
    \pgfusepath{fill}
    \pgfpathmoveto{\pgfpoint{0cm}{\i cm}}
    \pgfpathlineto{\pgfpoint{1cm - \i cm}{1cm}}
    \pgfpathlineto{\pgfpoint{1cm - \i cm - 0.125cm}{1cm}}
    \pgfpathlineto{\pgfpoint{0cm}{\i cm + 0.125cm}}
    \pgfpathclose
    \pgfusepath{fill}
  }
}
\tikzstyle{uncertain}=[circle,
                                pattern=stripes,
                                pattern color=gray!40,
                                thick,
                                minimum size=1.0cm,
                                draw=gray!80]
\tikzstyle{utility}=[regular polygon,regular polygon sides=6,
                                thick,
                                minimum size=1.0cm,
                                draw=gray!80,
                                fill=gray!20]
\tikzstyle{defensor_utility}=[regular polygon,regular polygon sides=6,
                                thick,
                                minimum size=1.0cm,
                                draw=gray!80,
                                fill=white]
\tikzstyle{decision}=[rectangle,
                                thick,
                                minimum size=1cm,
                                draw=gray!80,
                                fill=gray!20]
\tikzstyle{defensor_decision}=[rectangle,
                                thick,
                                minimum size=1cm,
                                draw=gray!80,
                                fill=white]

\tikzstyle{texto}=[label]
\begin{document}
\title{Gradient Methods for Solving Stackelberg Games\thanks{R.N. acknowledges the Spanish Ministry for his grant FPU15-03636. The work of D.R.I. is supported by the Spanish Ministry program MTM2017-86875-C3-1-R and the AXA-ICMAT Chair on Adversarial Risk Analysis. D.R.I also acknowledges the support of the EU's Horizon 2020 project 815003-2 TRUSTONOMY and the RTC-2017-6593-7 project.}}
%
%
\author{Roi Naveiro \and
David R\'ios Insua }
\authorrunning{R. Naveiro and D. R\'ios Insua}
%
\institute{Institute of Mathematical Sciences (ICMAT-CSIC), Madrid, Spain.}

\maketitle              
\begin{abstract}
Stackelberg Games are gaining importance in the last years due to the raise of Adversarial Machine Learning (AML). Within this context, a new paradigm must be faced: in classical game theory, intervening agents were humans whose decisions are generally discrete and low dimensional. In AML, decisions are made by algorithms and are usually continuous and high dimensional, e.g. choosing the weights of a neural network. As closed form solutions for Stackelberg games generally do not exist, it is mandatory to have efficient algorithms to search for numerical solutions. We study two different procedures for solving this type of games using gradient methods. We study time and space scalability of both approaches and discuss in which situation it is more appropriate to use each of them. Finally, we illustrate their use in an adversarial prediction problem.

\end{abstract}
\section{Introduction}
\label{sec:intro}

Over the last decade, the introduction of machine learning applications in numerous fields has grown tremendously. In particular, applications in security settings have grown substantially, \cite{mcdaniel2016machine}. In this domain, it is frequently the case that the data distribution at application time is different of the training data distribution, thus violating one of the key assumptions in machine learning. This difference between training and test distributions generally comes from the presence of adaptive adversaries who deliberately manipulate data to avoid being detected. 

The field of Adversarial Machine Learning (AML) studies, among other things, how to guarantee the security of machine learning algorithms against adversarial perturbations \cite{biggio2018wild}. A possible approach consists of modelling the interaction between the learning algorithm and the adversary as a game in which one agent controls the predictive model parameters while the other manipulates input data. Several different game theoretic models of this problem have been proposed, as reviewed in \cite{voro2018}. In particular, \cite{bruckner2011stackelberg} view adversarial learning as a Stackelberg game in which, a \textit{leader} (she), the defender in the security jargon, makes her decision about choosing the parameters in a learning model, and, then, the \textit{follower} or attacker (he), after having observed the leader's decision, chooses an optimal data transformation.

Mathematically, finding Nash equilibria of such Stackelberg games requires solving a bilevel optimization problem, which, in general cannot be undertaken analytically, \cite{sinha2018review}, and numerical approaches are required. However, standard techniques are not able to deal with continuous and high dimensional decision spaces, as those appearing in AML applications. 

In this paper, we propose two procedures to solve Stackelberg games in the new paradigm of AML and study their time and space scalability. In particular, one of the proposed solutions scales efficiently in time with the dimension of the decision space, at the cost of more memory requirements. The other scales well in space, but requires more time. The paper is organized as follows: in Section \ref{sec:stack_games} we define Stackelberg games. Section \ref{sec:solution_method} presents the proposed solution methods as well as a discussion of the scalability of both approaches. The proposed solutions are illustrated with an AML experiment in Section \ref{sec:experiments}. Finally, we conclude and present some lines for future research.

\section{Stackelberg games}
\label{sec:stack_games}

We consider a class of sequential games between two agents: the first one makes her decision, and then, after having observed the decision, the second one implements his response. These games have received various names in the literature including sequential Defend-Attack \cite{Brown:2006} or Stackelberg \citep{Gibbons:1992, tambe2011security} games. As an example, consider adversarial prediction problems, \cite{bruckner2011stackelberg}. In them, the first agent chooses the parameters of a certain predictive model; the second agent, after having observed such parameters, chooses an optimal data transformation to fool the first agent as much as possible, so as to obtain some benefit.

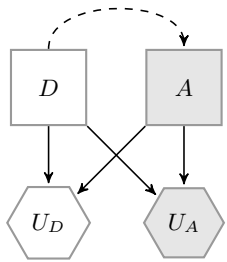
\begin{figure}[htbp]
\centering
\begin{tikzpicture}[->,>=stealth',shorten >=1pt,auto,node distance=1.8cm,
                    semithick]
  
  \node[defensor_decision] (A)   {$D$};
  \node[decision] (B) [right of=A] {$A$};
  \node[defensor_utility]  (C) [below of=A] {$U_D$};
  \node[utility]  (E) [below of=B] {$U_A$};

  \path (B) edge    node {} (C)
            edge    node {} (E)
        (A) edge    node {} (C)
            edge    node {} (E)
            edge[out =   90, in =  90, dashed]    node {} (B);
\end{tikzpicture}
 \caption{The two-player sequential decision game with certain outcome.} \label{fig:baid1}
\end{figure}

As we focus on applications of Stackelberg games to AML, we restrict ourselves to the
case in which the Defender ($D$) chooses her defense $\alpha \in \mathbb{R}^n$ and, then, the Attacker ($A$) chooses his attack $\beta \in \mathbb{R}^m$, after having observed $\alpha$. The corresponding bi-agent influence diagram, \cite{BAIDS}, is shown in Fig.~\ref{fig:baid1}. The dashed arc between nodes $D$ and $A$ reflects that the Defender choice is observed by the Attacker. The utility function of the Defender, $u_D(\alpha, \beta)$, depends on both, her decision, and the attacker's decision. Similarly, the Attacker's utility function has the form $u_A(\alpha, \beta)$. In this type of games, it is assumed that the Defender knows $u_A(\alpha, \beta)$. This assumption is known as the common knowledge hypothesis. 

Mathematically, finding Nash equilibrium of Stackelberg games requires solving a bilevel optimization problem, \cite{bard1991some}. The defender's utility is called \textit{upper level} or \textit{outer} objective function while the attacker's one is referred to as \textit{lower level} or \textit{inner} objective function. Similarly, the upper and lower level optimization problems, correspond to the defender's and the attacker's problem, respectively. These problems are also referred to as outer and inner problems.

It is generally assumed that the attacker will act rationally in the sense that he will choose an action that maximizes his utility, \cite{french2000statistical}, given the disclosed defender's decision $\alpha$. Assuming that there is a unique global maximum of the attacker's utility for each $\alpha$, and calling it $\beta^*(\alpha)$, a Stackelberg equilibrium is identified using backward induction: the defender has to choose $\alpha^*$ that maximizes her utility subject to the attacker's response $\beta^*(\alpha)$. Mathematically, the problem to be solved by the defender is 
\begin{equation}\label{bilevel}
\begin{aligned}
& \argmax_{\alpha}
& & u_D[\alpha, \beta^*(\alpha)]\\
& \text{s.t.}
& & \beta^*(\alpha) = \argmax_{\beta} u_A(\alpha,\beta).
\end{aligned}
\end{equation}
The pair $\left( \alpha^\opt , \beta^* (\alpha^\opt) \right)$ is a Nash equilibrium and a sub-game perfect equilibrium \citep{Hargreaves:2004}. 

When the attacker problem has more than one global maximum, several types of equilibrium have been proposed. The two more important are the optimistic and the pessimistic solutions, \cite{sinha2018review}. In an optimistic position, the defender expects the attacker to choose the optimal solution which gives the higher upper level utility. On the other hand, the pessimistic approach suggests that the defender should optimize for the worst case attacker solution. In this paper, we just deal with the case in which the inner utility has a unique global maximum.   

\section{Solution Method}
\label{sec:solution_method}

Bilevel optimization problems can rarely be solved analytically. Indeed even extremely simple instances of bilevel problems have been shown to be NP-hard, \cite{jeroslow1985polynomial}. Thus, numerical techniques are required. Several classical and evolutionary approaches have been proposed to solve \eqref{bilevel}, as reviewed by \cite{sinha2018review}. When the inner problem adheres to certain regularity conditions, it is possible to reduce the bilevel optimization problem to a single level one replacing the inner problem with its Karush-Kuhn-Tucker (KKT) conditions. Then, evolutionary techniques could be used to solve this single-level problem, thus making possible to relax the upper level requirements. As, in general, this single-level reduction is not feasible, several other approaches have been proposed, such as nested evolutionary algorithms or metamodeling-based methods. However, most of these approaches lack scalability: increasing the number of upper level variables produces an exponential increase on the number of lower level tasks required to be solved being thus impossible to apply these techniques to solve large scale bilevel problems as the ones appearing in the context of AML.

In \cite{bruckner2011stackelberg} the authors face the problem of solving Stackelberg games in the AML context. However, they focus on a very particular type of game which can be reformulated as a quadratic program. In this paper, we provide more general procedures to solve Stackelberg games that are useful in the AML paradigm in which decision spaces are continuous and high dimensional. To this end, we focus on gradient ascent techniques to solve bilevel optimization problems.

Let us assume that for any $\alpha$ the solution of the inner problem is unique. This solution defines an implicit function $\beta^*(\alpha)$. Thus, problem \eqref{bilevel} may be viewed solely in terms of the defender decisions $\alpha$. The underlying idea behind gradient ascent techniques is the following: given a defender decision $\alpha \in \mathbb{R}^n$ a direction along which the defender's utility increases while maintaining feasibility must be found, and then, we move $\alpha$ in that direction. Thus, the major issue of ascent methods is to find the gradient of $u_D(\alpha, \beta^*(\alpha))$. In \cite{kolstad1990derivative} the authors provide a method to approximate such gradient that work for relatively large classical optimization problems but it is clearly insufficient to deal with the typical bilevel problems appearing in AML.

Recently, \cite{franceschi2017forward} proposed forward and reverse-based methods for computing the gradient of the validation error in certain hyperparamenter optimization problems that appear in Deep Learning. Structurally, hyperparameter optimization problems are similar to Stackelberg games. We adapt their methodology to this domain. In particular we propose two alternative approaches to compute the gradient of $u_D[\alpha, \beta^*(\alpha)]$ with different memory and running time requirements. We refer to these approaches as backward and forward solutions, respectively.

\subsubsection*{Notation} 
For the sake of clarity, we use the following notation: the gradient will be denoted as $\dd_x$; the partial derivative as $\partial_x$. Similarly, second partial derivatives will be denoted as $\partial^2_x$ and $\partial_x \partial_y$. We shall use this notation indistinctly for the unidimensional and multidimensional cases. For instance, if $f(x,y)$ is a scalar function, $x$ is a $p$-dimensional vector and $y$ is a $q$-dimensional vector, then $\partial^2_x f(x,y)$ is the $p \times p$ matrix whose $(i,j)$ entry is $\partial_{x_i} \partial_{x_j} f(x,y)$, where $x_i$ is the $i$-th component of the vector $x$. Similarly, $\partial_x \partial_y f(x,y)$ is a $p \times q$ matrix whose $i,j$ entry is $\partial_{x_i} \partial_{y_j} f(x,y)$.

\subsection{Backward solution} \label{sec:backward}
We propose here a new gradient ascent approach to solve the bilevel problem \eqref{bilevel} whose running time scales well with the defender's decision space dimension. In particular, we propose to approximate problem \eqref{bilevel} by the following PDE-constrained optimization problem, \cite{hinze2008optimization}
\begin{equation}\label{alternative}
\begin{aligned}
& \argmax_{\alpha}
& & u_D\left[\alpha, \beta(\alpha, T)\right] \\
& \text{s.t.}
& & \partial_t \beta(\alpha, t) = \partial_{\beta} u_A [\alpha,  \beta(\alpha, t)]\\
& & & \beta(\alpha, 0) = 0.
\end{aligned}
\end{equation}
The idea is formalized in the next proposition, that can be proved using the results in \cite{bottou1998online}.
\begin{proposition}
Suppose that the following assumptions hold
\begin{enumerate}
    \item The attacker problem, the inner problem in \eqref{bilevel}, has a unique solution $\beta^*(\alpha)$ for each defender decision $\alpha$.
    \item For all $\epsilon>0$ and all $\alpha$,
    \begin{eqnarray*}
     \inf_{\Vert \beta^*(\alpha)  - \beta  \Vert_2^2 > \epsilon} \left \langle \beta - \beta^*(\alpha), \partial_{\beta}u_A [\alpha, \beta] \right \rangle > 0.   
    \end{eqnarray*}
\end{enumerate}    
If $\beta(\alpha, t)$ satisfies the differential equation
\begin{equation} \label{diffeq}
         \partial_t \beta(\alpha, t) = \partial_{\beta} u_A [\alpha, \beta(\alpha, t)]
\end{equation}
then $\beta(\alpha, t) \rightarrow \beta^*(\alpha)$ as $t \rightarrow \infty$, with rate $\mathcal{O}\left(\frac{1}{t}\right)$.
\end{proposition}
\noindent The idea in $\eqref{alternative}$ is thus to constrain the trajectories $\beta(\alpha, t)$ to satisfy $\eqref{diffeq}$ and approximate the defender's problem using $\beta(\alpha, T)$ with $T \gg 1$, instead of $\beta^*(\alpha)$.

We propose solving problem \eqref{alternative} using gradient ascent and the adjoint method, \cite{pontryagin2018mathematical}, to compute the total derivative of the defender utility function with respect to her decision. The adjoint method defines an \textit{adjoint function} $\lambda(t)$ satisfying the \textit{adjoint equation}
\begin{equation} \label{adjointeq}
        \dd_t \lambda (t) = - \lambda(t) ~ \partial_{\beta}^2 u_A[\alpha, \beta(\alpha, t)].
\end{equation}
In terms of the adjoint function, the derivative of the defender utility with respect to her decision would be written as
\begin{equation} \label{derivative}
        \dd_\alpha u_D[\alpha, \beta(\alpha, T)] = \partial_\alpha u_D[\alpha, \beta(\alpha, T)] - \int_0^T \lambda(t) \partial_{\alpha} \partial_{\beta} u_A [\alpha, \beta(\alpha, t)] \dd t.
\end{equation}
In Appendix \ref{prf}, we prove that if $\lambda(t)$ satisfies the adjoint equation \eqref{adjointeq}, the derivative of the defender utility can be written as in \eqref{derivative}.

Algorithmically, we can proceed by discretizing \eqref{adjointeq} via Euler method, and approximate the derivative \eqref{derivative} discretizing the integral on the left hand side. This leads to Algorithm \ref{alg:aprox_der_bw}.
\begin{algorithm} 
  \caption{Approximate total derivative of defender utility function with respect to her decision using the backward solution} \label{alg:aprox_der_bw}
  \begin{algorithmic}[1]
    \Procedure{Approximate Derivative using Backward Method}{$\alpha, T$}
    \State $\beta_0(\alpha) = 0$
      \For{$t = 1,2,\dots,T$} 
        \State $\beta_t(\alpha) = \beta_{t-1}(\alpha) + \eta \partial_{\beta}u_A(\alpha, \beta)\Big \vert_{\beta_{t-1}}$
    \EndFor
      \State $\lambda_T = - \partial_{\beta} u_D(\alpha, \beta)\Big \vert_{\beta_{T}}$
      \State $\dd_{\alpha} u_D = \partial_{\alpha} u_D[\alpha, \beta_T(\alpha)]$
       \For{$t = T-1, T-2, \dots, 0$} 
        \State $\dd_{\alpha} u_D = \dd_{\alpha} u_D - \eta \lambda_{t+1} \partial_\alpha \partial_\beta u_A(\alpha, \beta)\Big \vert_{\beta_{t+1}}$
        \State $\lambda_t = \lambda_{t+1} \left[I + \eta \partial^2_{\beta}u_A(\alpha, \beta)\Big \vert_{\beta_{t+1}} \right]$
    \EndFor
    \State \textbf{return} $\dd_{\alpha} u_D$
    \EndProcedure
  \end{algorithmic}
\end{algorithm}
\noindent Once we are able to compute this derivative, we can solve the defender's problem using gradient ascent.

Regarding its complexity, note that by basic facts of Automatic Differentiation (AD), \cite{griewank2008evaluating}, if $\tau (n,m)$ is the time required to evaluate $u_D(\alpha, \beta)$ and $u_A(\alpha, \beta)$, then computing derivatives of these functions requires time $\mathcal{O}(\tau (n,m))$. Thus the first for loop in Algorithm \ref{alg:aprox_der_bw} requires time $\mathcal{O}(T \tau (n,m))$. In the second loop, we need to compute second derivatives, which appear always multiplying the vector $\lambda_t$. By basic results of AD, Hessian vector products have the same time complexity as function evaluations. Thus in our case, we can compute second derivatives in time $\mathcal{O}(\tau (n,m))$ being the time complexity of the second for loop $\mathcal{O}(T\tau (n,m))$. Thus, overall, Algorithm \ref{alg:aprox_der_bw} runs in time $\mathcal{O}(T\tau (n,m))$. Regarding space complexity, as it is necessary to store the values of $\beta_t(\alpha)$ produced in the first loop for later usage in the second one, if $\sigma(n,m)$ is the space requirement for storing each $\beta_t(\alpha)$, then $\mathcal{O}(T \sigma(n,m))$ is the space complexity of the backward algorithm.

In certain applications where space complexity is critical, the backward solution could be infeasible as, within each iteration, it requires storing the whole trace $\beta_t(\alpha)$. In this particular cases, the forward solution proposed in the next section, solves this issue at a cost of loosing time scalability.

\subsection{Forward solution}

In this case, we approximate \eqref{bilevel} by
\begin{equation}\label{alternative_fw}
\begin{aligned}
& \argmax_{\alpha}
& & u_D\left[\alpha, \beta_T(\alpha)\right] \\
& \text{s.t}
& & \beta_t(\alpha) = \beta_{t-1}(\alpha) + \eta_t \partial_{\beta}u_A(\alpha, \beta)\Big \vert_{\beta_{t-1}} && t = 1, \dots, T\\
& & & \beta_0(\alpha) = 0.
\end{aligned}
\end{equation}
The idea here is that, for each defense $\alpha$, we condition on a dynamical system that under certain conditions converges to $\beta^*(\alpha)$, the optimal solution for the attacker when the defender plays $\alpha$. Thus, we can approximate the defender's utility by $u_D\left[\alpha, \beta_T(\alpha)\right]$, with $T \gg 1$. This idea is formalized in the next proposition that can be proved using the results of \cite{bottou1998online}.
\begin{proposition}
Suppose that the following assumptions hold

\begin{enumerate}
    \item The attacker problem (the inner problem in \eqref{bilevel}) has a unique solution $\beta^*(\alpha)$ for each defender decision $\alpha$.
    \item For all $\epsilon>0$ and $\alpha$
    \begin{equation*}
        \inf_{\Vert \beta - \beta^*(\alpha) \Vert_2^2 > \epsilon} \left \langle \beta - \beta^*(\alpha), \partial_{\beta}u_A [\alpha, \beta] \right \rangle > 0
    \end{equation*}
    \item For some $A,B>0$ and all $\alpha$
    \begin{equation*}
        \Vert \partial_{\beta}u_A [\alpha, \beta] \Vert_{2}^2 \leq A + B \Vert \beta - \beta^*(\alpha) \Vert_{2}^2
    \end{equation*}
\end{enumerate}    
If  for all $t$, $\beta_t(\alpha)$ satisfies
\begin{equation} \label{dynsys}
         \beta_t(\alpha) = \beta_{t-1}(\alpha) + \eta \partial_{\beta}u_A(\alpha, \beta)\Big \vert_{\beta_{t-1}}
\end{equation}
Then, $\beta_t(\alpha)$ converges to $\beta^*(\alpha)$, with rate $\mathcal{O}\left(\frac{1}{t}\right)$.
\end{proposition}

We propose solving problem \eqref{alternative_fw} using gradient ascent. To that end, we need to compute $\dd_\alpha u_D(\alpha, \beta_T(\alpha))$. Using the chain rule we have
\begin{eqnarray*}
    \dd_\alpha u_D[\alpha, \beta_T(\alpha)] = \partial_\alpha u_D[\alpha, \beta_T(\alpha)] + \partial_{\beta_T} u_D[\alpha, \beta_T(\alpha)] \dd_\alpha \beta_T(\alpha)
\end{eqnarray*}
To obtain $\dd_\alpha \beta_T(\alpha)$, we can sequentially compute $\dd_\alpha \beta_t(\alpha)$ taking derivatives in \eqref{dynsys}
\begin{eqnarray*}
\dd_\alpha \beta_t(\alpha) = \dd_\alpha \beta_{t-1}(\alpha) + \eta_{t-1} \left[ \partial_\alpha \partial_{\beta} u_A(\alpha, \beta)\Big \vert_{\beta_{t-1}} + \partial^2_\beta u_A(\alpha, \beta)\Big \vert_{\beta_{t-1}} \dd_\alpha \beta_{t-1}(\alpha) \right]
\end{eqnarray*}
This induces a dynamical system in $\dd_\alpha \beta_t(\alpha)$ that can be iterated in parallel to the dynamical system in $\beta_t(\alpha)$. The whole procedure is described in Algorithm \ref{alg:aprox_der_fw}.
\begin{algorithm}
  \caption{Approximate total derivative of defender utility function with respect to her decision using the forward solution.} \label{alg:aprox_der_fw}
  \begin{algorithmic}[1]
    \Procedure{Approximate Derivative using Forward Method}{$\alpha, T$}
    \State $\beta_0(\alpha) = 0$
    \State $\dd_\alpha \beta_0(\alpha) = 0$
      \For{$t = 1,2,\dots,T$} 
        \State $\beta_t(\alpha) = \beta_{t-1}(\alpha) + \eta \partial_{\beta}u_A(\alpha, \beta)\Big \vert_{\beta_{t-1}}$
        \State $\dd_\alpha \beta_t(\alpha) = \dd_\alpha \beta_{t-1}(\alpha) + \eta_{t-1} \left[ \partial_\alpha \partial_{\beta} u_A(\alpha, \beta)\Big \vert_{\beta_{t-1}} + \partial^2_\beta u_A(\alpha, \beta)\Big \vert_{\beta_{t-1}} \dd_\alpha \beta_{t-1}(\alpha) \right]$
    \EndFor
      \State $\dd_\alpha u_D = \partial_\alpha u_D[\alpha, \beta_T(\alpha)] + \partial_{\beta_T} u_D[\alpha, \beta_T(\alpha)] \dd_\alpha \beta_T(\alpha)$
    \State \textbf{return} $\dd_{\alpha} u_D$
    \EndProcedure
  \end{algorithmic}
\end{algorithm}
Once we are able to compute this derivative, we can solve the defender's problem using gradient ascent.

Regarding time complexity, note that the bottleneck in Algorithm \ref{alg:aprox_der_fw} is that we need to compute second derivatives of $u_A(\alpha,\beta)$. In particular, computing $\partial^2_\beta u_A(\alpha, \beta)$ requires time $\mathcal{O}(m \tau(m,n))$ as it requires computing $m$ Hessian vector products, one with each of the $m$ the unitary vectors. On the other hand, computing $\partial_\alpha \partial_{\beta} u_A(\alpha, \beta)$ requires computing $n$ Hessian vector products and thus time $\mathcal{O}(n \tau(m,n))$, while if we compute the derivative in the other way, first we derive with respect to $\beta$ and then with respect to $\alpha$, the time complexity is $\mathcal{O}(m \tau(m,n))$. Thus, we derive first with respect to the variable with the biggest dimension. Then, the time complexity of computing $\partial_\alpha \partial_{\beta} u_A(\alpha, \beta)$ is $\mathcal{O}(\min(n,m) \tau(m,n))$. Finally, as $\partial^2_\beta u_A(\alpha, \beta)$  and $\partial_\alpha \partial_{\beta} u_A(\alpha, \beta)$ could be computed in parallel, then the overall time complexity of the forward solution is $\mathcal{O}(\max[\min(n,m), m]T \tau(m,n)) = \mathcal{O}(mT \tau(m,n)) $. Regarding space, as in this case the values $\beta_t(\alpha)$ are overwritten at each iteration, we do not need to store all of them and the overall space complexity is $\mathcal{O}( \sigma(m,n))$.

\section{Experiments}
\label{sec:experiments}
We illustrate now the proposed approaches. We start with a conceptual example in which we empirically test the scalability properties of both algorithms. Then, we apply the algorithms to solve a problem in the context of adversarial regression. 

All the code used for these examples has been written in python using the pytorch library for Automatic Differentiation, \cite{paszke2017automatic}; and is available at \url{https://github.com/roinaveiro/GM_SG}.

\subsection{Conceptual Example}
We use a simple example to illustrate the scalability of the proposed approaches. Consider that the attacker's and defender's decisions are both vectors in $\mathbb{R}^n$. The attacker's utility takes the form $u_A(\alpha, \beta) = -\sum_{i=1}^n 3(\beta_i - \alpha_j)^2$ and the defender's one is $u_D(\alpha, \beta) = -\sum_{i=1}^n (7 \alpha_i + \beta_j^2)$. In this case, the equilibrium can be computed analytically using backward induction: for a given defense $\alpha \in \mathbb{R}^n$ we see that $\beta^*(\alpha) = \alpha$; substituting in the outer problem, the equilibrium is reached at $\alpha_j^* = -3.5, \beta_j^*(\alpha^*) = -3.5$ with $j=1, \dots, m$. 

We apply the proposed methods to this problem to test their scalability empirically. The parameters were chosen as follows: the learning rate $\eta$ of Algorithms \ref{alg:aprox_der_bw} and \ref{alg:aprox_der_fw} was set to 0.1; similarly, the learning rate of the gradient ascent used to solve the outer problem was also set to 0.1. Finally, all gradient ascents were run for $T=40$, enough to reach convergence.
\begin{figure}[h!]
 \centering
\includegraphics[width=0.8\linewidth]{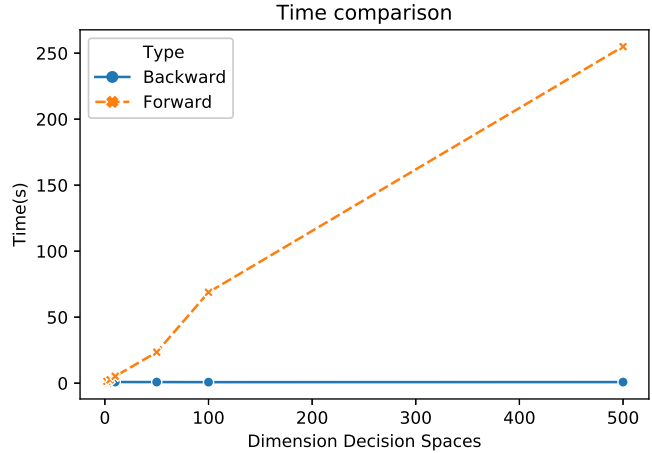}
\caption{Backward and Forward running times versus the dimension of decision spaces.} \label{fig:time_comp}
\end{figure}    

Figure \ref{fig:time_comp} shows running times for increasing number of dimensions of the decision spaces (in this problem both the attacker's and the defender's decision space have the same dimension). As we discussed, the forward running time increases linearly with the number of dimensions while the backward solution remains approximately constant. This obviously comes at the cost of having more memory requirements, as in Algorithm \ref{alg:aprox_der_bw} we need to store the whole trace $\beta_t(\alpha)$. Thus, in problems where the dimension of $\beta$ is very large the memory cost of the backward solution would become prohibitive and we would need to switch to the forward solution, as long as the dimension of $\alpha$ is small enough. In contrast, if the dimension of $\alpha$ is very big, the forward solution would become infeasible in time, thus being the backward optimal provided that the dimension of $\beta$ is such that it is possible to store the whole trace $\beta_t(\alpha)$.

\subsection{An application to adversarial regression} \label{exp:adv}

\subsubsection{Problem statement}

We illustrate an application of the proposed methodology to adversarial regression problems, \cite{grosshans2013bayesian}. They are a specific class of prediction games, \cite{bruckner2011stackelberg}, played between a \textit{learner} of a regression model and a \textit{data generator}, who tries to fool the learner modifying input data at application time, inducing a change between the data distribution at training and test time, with the aim of confusing the data generator and attain a benefit.

Given a feature vector $x \in \mathbb{R}^p$ and its corresponding target value $y \in \mathbb{R}$, the learner's decision is to choose the weight vector $w \in \mathbb{R}^p$ of a linear model $f_w(x) = x^\top w$, that minimizes the theoretical costs at application time, given by
\begin{eqnarray*}
\theta_l (w, \bar{p}, c_l) = \int c_l(x,y) (f_w(x) - y)^2 \dd \bar{p}(x,y),
\end{eqnarray*}
where $c_l(x,y) \in \mathbb{R}^+$ reflects instance-specific costs and $\bar{p}(x,y)$ is the data distribution at test time. To do so, the learner has a training matrix $X \in \mathbb{R}^{n\times p}$ and a vector of target values $y\in \mathbb{R}^n$, that is a sample from distribution $p(x,y)$ at training time.

The data generator aims at changing features of test instances to induce a transformation in the data distribution from $p(x,y)$ to $\bar{p}(x,y)$. Let $z(x,y)$ be the data generator's target value for instance $x$ with real value $y$, i.e.\ he aims at transforming $x$ to make the learner predict $z(x,y)$ instead of $y$. The data generator aims at choosing the transformation that minimizes the theoretical costs given by
\begin{eqnarray*}
\theta_d (w, \bar{p}, c_d) = \int c_d(x,y) (f_w(x) - z(x,y))^2 \dd \bar{p}(x,y) + \Omega_d(p,\bar{p})
\end{eqnarray*}
where $\Omega_d(p,\bar{p})$ is the incurred cost when transforming $p$ to $\bar{p}$ and $c_d(x,y)$ are instance specific costs. 

As the theoretical costs defined above depend on the unknown distributions $p$ and $\bar{p}$, we focus on their regularized empirical counterparts, given by
\begin{eqnarray*}
\widehat{\theta}_l(w, \bar{X}, c_l) &=& \sum_{i=1}^n c_{l,i} (f_w(\bar{x}_i) - y_i)^2 + \Omega_l(f_w),\\
\widehat{\theta}_d(w, \bar{X}, c_d) &=& \sum_{i=1}^n c_{d,i} (f_w(\bar{x}_i) - z_i)^2 + \Omega_d(X, \bar{X}).
\end{eqnarray*}
In addition, we assume that the learner acts first, choosing a weight vector $w$. Then the data generator, after observing $w$, chooses his optimal data transformation. Thus, the problem to be solved by the learner is
\begin{equation} \label{advstack}
\begin{aligned}
& \argmin_{w}
& & \widehat{\theta}_l(w, T(X,w,c_d), c_l)  \\
& \text{s.t.}
& & T(X,w,c_d) = \argmin_{X'} \widehat{\theta}_d(w, X', c_d), 
\end{aligned}
\end{equation}
where $T(X,w,c_d)$ is the attacker's optimal transformation for a given choice $w$ of weight vector. \eqref{advstack} has the same form as \eqref{bilevel}, except that it is formulated in terms of costs rather than utilities. In addition, it is easy to see that if $\Omega_d(X, \bar{X})$ is equal to the squared Frobenius norm of the difference matrix $\Vert X - \bar{X} \Vert_F^2$, then the attacker's problem has a unique solution. Thus, we can use the proposed solution techniques to look for Nash equilibria in this type of game, taking care of performing gradient descent instead of gradient ascent, as we are minimizing costs here.

\subsubsection{Experimental results}
We apply the results to the UCI white wine dataset, \cite{UCI}. This contains real information about 4898 wines, that consists of 11 quality indicators plus a wine quality score. 

$R_J$ and $R_D$ are two competing wine brands. $R_D$ has implemented a system to automatically measure wine quality using a regression over the available quality indicators: each wine is described by a vector of 11 entries, one per quality indicator. Wine quality ranges between 0 and 10. $R_J$, aware of the actual superiority of its competitor's wines, decides to hack $R_D$'s system by manipulating the value of several quality indicators, to artificially decrease $R_D$'s quality rates. However, $R_D$ is aware of the possibility of being hacked, and decides to use adversarial methods to train its system. In particular, $R_D$ models the situation as a Stackelberg game. It is obvious that the target value of his enemy is $z(x,y) = 0$ for every possible wine. In addition, $R_D$ was able to filter some information about $R_J$'s wine-specific costs $c_{d,i}$.

As basic model, a regular ridge regression, \cite{friedman2001elements}, was trained using eleven principal components as features. The regularization strength was chosen using repeated hold-out validation, \cite{kim2009estimating}, with ten repetitions. As performance metric we used the root mean squared error (RMSE), estimated via repeated hold-out.

We compare the performance of two different learners against an adversary whose wine specific costs $c_{d,i}$ are fixed: The first one, referred to as Nash, assumes that the wine specific costs are common knowledge and plays Nash equilibria of the Stackelberg game defined in \eqref{advstack}. The second learner, refer to as raw, is a non adversarial one and uses a ridge regression model. To this end, we split the data in two parts, $2/3$ for training purposes and the remaining $1/3$ for test. The training set is used to compute the weights $w$ of the regression problem. Those weights are observed by the adversary, and used to attack the test set. Then, the RMSE is computed using this attacked test set and the previously computed weights.

In order to solve \eqref{advstack}, we use the backward solution solution method of Section \ref{sec:backward} due to its better time scalability. The hyperparameters were chosen as follows: number of epochs $T$ to compute the gradient in Algorithm \ref{alg:aprox_der_bw}, 100; the learning rate $\eta$ in this same Algorithm, was set to 0.01. Within the gradient descent optimization used to optimize the defender's cost function, the number of epochs was set to 350 and the learning rate to $10^{-6}$. Finally, we assumed that the wine specific costs were the same for all instances and called the common value $c_d$. We studied how $c_d$ affects the RMSE for different solutions. 

Notice that, in this case, the dimension of the attacker's decision space is huge. He has to modify the training data to minimize his costs. If there are $k$ instances in the training set, each of dimension $n$, the dimension of the attacker's decision space is $n\times k$. In this case $k=3263$ ($2/3$ of 4898) and $n=11$. Thus the forward solution is impractical in this case, and we did not compute it.
\begin{figure}[t!]%
\begin{minipage}{0.5\textwidth}%
\includegraphics[height=1.9in, width=1\linewidth]{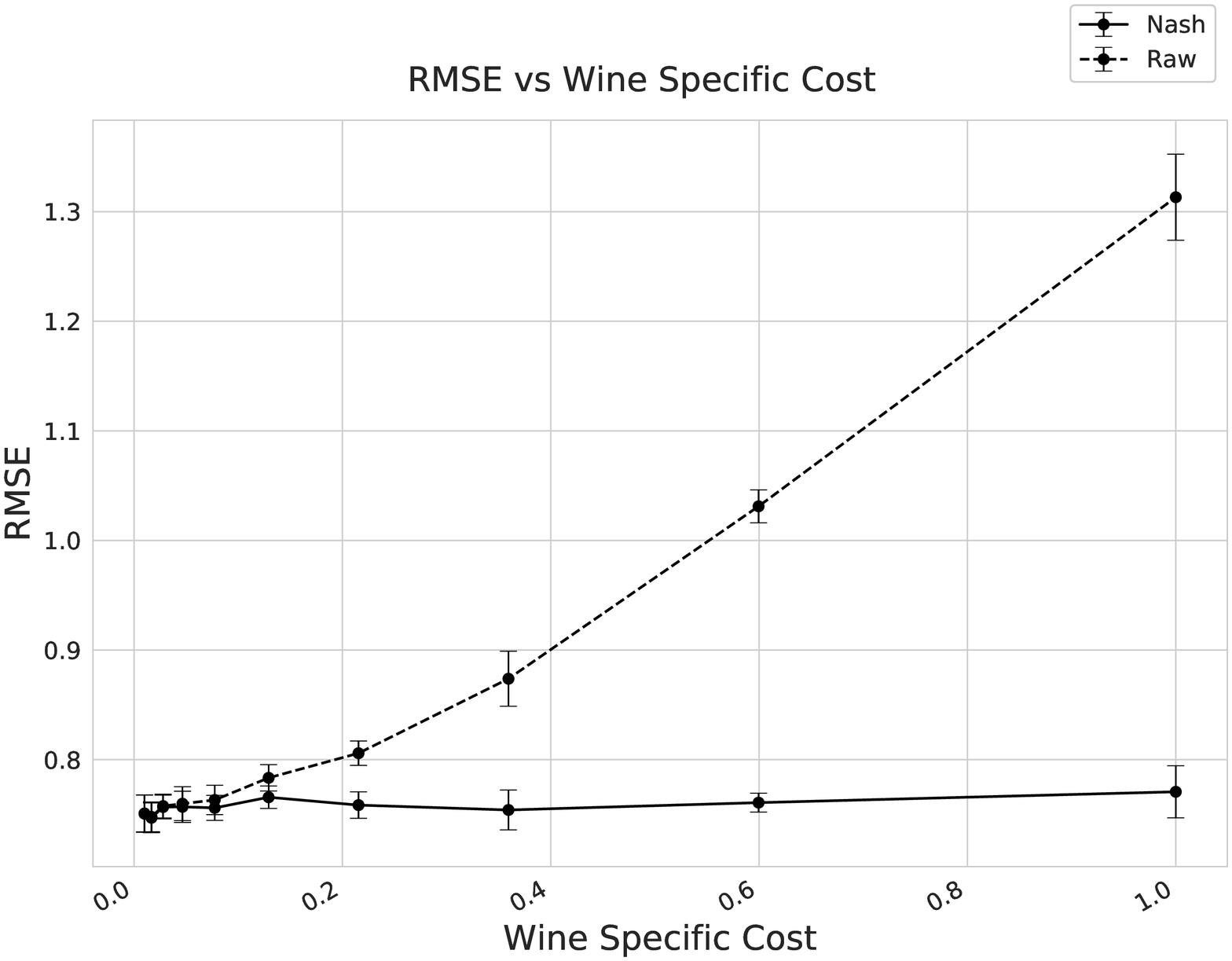}
\caption{Performance comparison.} \label{performance}
\end{minipage}%
\begin{minipage}{0.5\textwidth}%
\includegraphics[height=1.82in, width=0.95\linewidth]{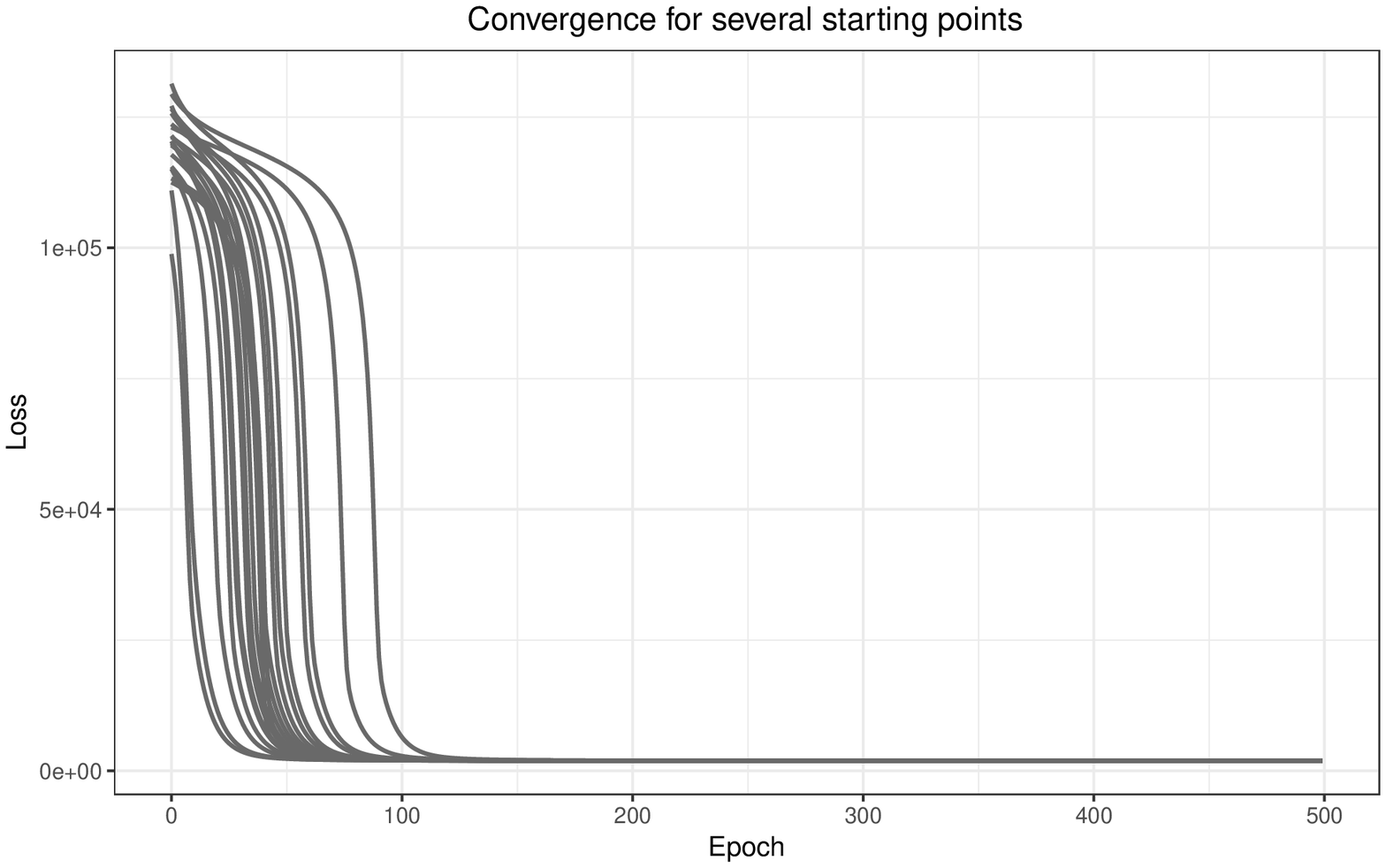}
\caption{Convergence for several initial points.} \label{convergence}
\end{minipage}%
\end{figure}
We show in Figure \ref{performance}, the RMSE for different values of the wine specific cost. We observe that Nash outperforms systematically the adversary unaware regression method. In the limit $c_d \rightarrow 0$, we see that $\widehat{\theta}_d(w, \bar{X}, c_d) \rightarrow \Omega_d(X, \bar{X})$. Thus, in this situation, the adversary will not manipulate the data. Consequently Nash and ridge regression solutions will coincide, as shown in Figure \ref{performance}. However, as $c_d$ increases, data manipulation is bigger, and the RMSE of the adversary unaware method also increases. On the other hand, the Nash solution RMSE remains almost constant.

We have also computed the average and standard deviation of training times. In an Intel Core i7-3630UM, 2.40GHz × 8 computer, the average training time is $131.6$ seconds with $2.7$ seconds standard deviation. This corresponds approximately to $2.66$ seconds per outer epoch. Each outer epoch involves running Algorithm \ref{alg:aprox_der_bw} with 100 inner epochs.

Finally, to illustrate convergence of the proposed approach, we solve \eqref{advstack} using gradient descent with the backward method for 20 different random initializations of the defender's decisions $\omega$. Results are depicted in Figure \ref{convergence}. As can be seen, all paths converge with less than 150 epochs.

\section{Discussion}
\label{sec:discussion}

The demand for scalable solutions of Stackelberg Games has increased in the last years due to the use of such games to model confrontations within Adversarial Machine Learning problems. In this paper, we have focused on gradient methods for solving Stackelberg Games, providing two different approaches to compute the gradient of the defender's utility function: the forward and backward solutions. In particular, we have shown that the backward solution scales well in time with the defender's decision space dimension, at a cost of more memory requirements. On the other hand, the forward solution scales poorly in time with this dimension, but well in space.

We have provided empirical support of the scalability properties of both approaches using a simple example. In addition, we have solved an AML problem using the backward solution in a reasonable amount of time. In this problem, the defender's decision space is continuous with dimension 11. The attacker's decision space is also continuous with dimension $\mathcal{O}(10^4)$, as we showed in Section \ref{exp:adv}. To the best of our knowledge, none of previous numerical techniques for solving Stackelberg games could deal, in reasonable time, with such high dimensional continuous decision spaces. 

Apart from scalability properties, a major advantage of the proposed framework is that it could be directly implemented in any Automatic Differentiation library such as PyTorch (the one used in this example) or TensorFlow, and thus benefit from the computational advantages of such implementations.

We could extend the framework in several ways. First, as we discussed, the backward solution has poor space scalability. This is generally not an issue in most applications. Nevertheless, if space complexity is critical it is possible to reduce it at a cost of introducing a numerical error, as proposed in \cite{maclaurin2015gradient} in hyperparameter optimization problems. Instead of storing the whole trace $\beta_t(\alpha)$ in the first for loop of Algorithm \ref{alg:aprox_der_bw} to use it in the second loop, we could sequentially undo its gradient update at each step of the second for loop. Obviously, this would introduce some numerical error.

Another possible line of work would be to extend the framework to deal with Bayesian Stackelberg games, that are widely used to model situations in AML in which there is not common knowledge of the adversary's parameters. In this line, the ultimate goal would be to apply the proposed algorithms to solve Adversarial Risk Analyisis (ARA, \cite{rios2009adversarial}) problems in AML, \cite{naveiro2019adversarial}. 

Throughout the paper, we have focused on exact gradient methods. However, it would be interesting to extend the proposed algorithms to work with stochastic gradient methods. In addition, in \cite{mokhtari2019unified} the authors propose several variants of Gradient Ascent to solve saddle point problems. It could be worth investigating how to extend such techniques to general Stackelberg Games.

Finally, we highlight that one of the most important contributions of the paper is the derivation of the backward solution formulating the Stackelberg game \eqref{bilevel} as a PDE-constrained optimization problem and using the adjoint method. This provides a general and scalable framework that could be used to seek for Nash equilibria in other types of sequential games. Exploring this, is another possible line of future work. 
%
%
%
%
\vspace{1.5cm}

\bibliographystyle{splncs04}
{
\renewcommand{\clearpage}{} 
\bibliography{references}
}
%
\appendix

\section{Proof of the adjoint method} \label{prf}
The Lagrangian of problem \eqref{alternative} is
\begin{equation*}
    \mathcal{L} = u_D[\alpha, \beta(\alpha, T)] + \int_{0}^T \lambda(t) \left \lbrace \dd_t \beta(\alpha, t) - \partial_{\beta} u_A [\alpha, \beta(\alpha, t)] \right \rbrace dt + \mu \beta(\alpha, 0). 
\end{equation*}
As the constraints hold, by construction we have that $d_{\alpha}\mathcal{L} = d_{\alpha}u_D$ and
\begin{eqnarray}
    \dd_\alpha \mathcal{L} &=& \partial_\alpha u_D[\alpha, \beta(\alpha, T)] + \partial_\beta u_D[\alpha, \beta(\alpha, T)] \dd_\alpha \beta(\alpha, T) + \mu \dd_\alpha \beta(\alpha, 0) \nonumber \\ 
    &+& \int_0^T \lambda(t)  \left \lbrace \dd_t \dd_\alpha \beta(\alpha, t) - \partial_\alpha \partial_{\beta} u_A [\alpha, \beta(\alpha, t)] - \partial^2_{\beta} u_A [\alpha, \beta(\alpha, t)] \dd_\alpha \beta(\alpha, t) \right \rbrace \dd t. \nonumber \\ \label{first_adj}
\end{eqnarray}
Integrating by parts, we have
\begin{eqnarray*}
    \int_0^T \lambda(t) \dd_t \dd_\alpha \beta(\alpha, t)  \dd t = \left[\lambda(t)\dd_\alpha \beta(\alpha, t) \right]_0^T - \int_0^T \dd_t\lambda(t) \dd_\alpha \beta(\alpha, t) \dd t
\end{eqnarray*}
Inserting this in \eqref{first_adj} and grouping the terms conveniently we have
\begin{eqnarray*}
    \dd_\alpha \mathcal{L} &=& \partial_\alpha u_D[\alpha, \beta(\alpha, T)] + \Big\lbrace \partial_\beta u_D[\alpha, \beta(\alpha, T)] + \lambda(T) \Big\rbrace \dd_\alpha \beta(\alpha, T) + \lbrace\mu- \lambda(0) \rbrace \dd_\alpha \beta(\alpha, 0) \nonumber \\ 
    &+& \int_0^T \left \lbrace - \dd_t\lambda(t) - \lambda(t)\partial^2_{\beta} u_A [\alpha, \beta(\alpha, t)]\right \rbrace  \dd_\alpha \beta(\alpha, t) -\lambda(t) \partial_\alpha \partial_{\beta} u_A [\alpha, \beta(\alpha, t)]  \dd t
\end{eqnarray*}
Since the constraints hold, we may choose freely the Lagrange multipliers. In particular, we may choose them so that we can avoid calculating the derivatives of $\beta(\alpha, t)$ with respect to $\alpha$ (as this is computationally expensive). Thus, we have that $\lambda$ satisfies the adjoint equation
\begin{eqnarray*}
    \dd_t\lambda(t) =  - \lambda(t)\partial^2_{\beta} u_A [\alpha, \beta(\alpha, t)]
\end{eqnarray*}
with $\lambda(T) = - \partial_\beta u_D[\alpha, \beta(\alpha, T)]$, and $\mu = \lambda(0)$. Using this, the derivative is computed as

\begin{eqnarray*}
    \dd_\alpha \mathcal{L} &=& \partial_\alpha u_D[\alpha, \beta(\alpha, T)] - \int_0^T \lambda(t) \partial_\alpha \partial_{\beta} u_A [\alpha, \beta(\alpha, t)] \dd t
\end{eqnarray*}



\end{document}